\documentclass[12pt,a4paper]{article}

\usepackage[T2A]{fontenc}

\usepackage[maccyr]{inputenc}

\usepackage[russian,english]{babel}

\usepackage[dvips]{color}
\usepackage{epsfig,color}
\usepackage{pstricks,graphicx,epsfig,color,amssymb,amsmath,amscd}
\usepackage{bm}

\newcommand{\bi}{\bibitem}

\def\be{\begin{eqnarray}}
\def\ee{\end{eqnarray}}

\begin{document}

\title{\bf 
Massive and supermassive black holes in the contemporary and early universe
and the new problems of cosmology and astrophysics.}
\author{A.D. Dolgov }
\maketitle

\begin{center}
Novosibirsk State University, Novosibirsk 630090\\
Institute of Theoretical and Experimental Physics, Moscow 117218  
\end{center}

\begin{abstract}

This is the translation into English of the introduction, conclusion, and the list of references  of the review on massive 
primordial black holes, which is submitted in Russian to Uspekhi Fizicheskikh Nauk (Physics-Uspekhi). If accepted, 
this review is translated into English by the Journal and published in Russian and a little later in English.

The review  concerns  the recent astronomical data which show that massive primordial black holes play much more 
significant role in the universe than it was previously believed. This is true both for the the contemporary and the early 
universe at  the red-shifts about 10. The mechanism, proposed in 1993, of primordial creation of heavy and superheavy 
black holes in the very early universe is discussed. This mechanism predicts the log-normal mass spectrum of the primordial 
black holes, which became very popular during the last couple of years. The proposed mechanism presents a natural 
explanation of a large amount of the recent observational data, which do not fit the standard cosmology and astrophysics.

 \end{abstract}

{ \bf Content}\\
1. Introduction.\\
2. Universe population at $z \sim 10 $.\\
2.1. Universe age as a function of redshift.\\
2.2. Bright galaxies in the early universe.\\
2.3. Quasars and supermassive black holes at  $ z> 6$.\\
2.4. Dust, supernovae, and gamma-bursters.\\
3. Puzzles of the contemporary and near contemporary universe. \\
3.1. Old stars in the Milky Way. \\
3.2. Supermassive black holes at the present days.\\
3.3. (Near)Solar mass black holes.\\
3.4. Problems of MACHOs.\\
3.5. Intermediate mass black holes in the contemporary universe.\\
4. Problems with the sources of gravitational waves discovered by LIGO.\\
5. Early creation of heavy and superheavy black holes and compact stellar-like objects.\\
6. Rise of the PBH masses due to accretion.\\
6.1. Conventional approach.\\ 
6.2.Rise of baryonic bubbles.\\
7. Globular clusters.\\
8. Impact of PBH on BBN and CMB.\\
9. Electromagnetic radiation from gravitational waves.\\
10. Conclusion.\\
11. References.

\section{Introduction} \label{s-intro} 

Significant improvement of the telescope sensitivity (in all wave lengths) during the last few years has lead to a number
of surprising astronomical discoveries, especially  impressive in the universe at $z = 5-10$, which corresponds to the
universe age from 400 million up to a billion years.
It was found that such a young universe  was densely populated by the objects which, according to the accepted views,
simply could not appear there in such a short time.

In the universe at this early stage plenty of well evolved objects are discovered, 
including quasars (alias supermassive black holes) with the masses of
billions solar masses ($M_{\odot}$), very bright and huge galaxies, supernovae, and gamma-bursters. The universe was 
full of metals and quite dusty.

Similar problems manifest themselves in the present day universe as well. They became especially pronounced after the recent 
discoveries at high z. In any large elliptic or venticular galaxy there is a supermassive black hole (SMHB) with the mass of
about billion solar masses. In spiral galaxies, such as e.g. our Milky Way, the central black holes also live  but with much
smaller masses, of a few million solar masses. The canonical theories of such black hole formation, especially of the supermassive ones,
through  the matter accretion on the galactic center encounter serious problems so the existence of such black holes is one
of most serious astrophysical challenges.

In this connection a natural idea may come to one's mind, that the SMBH are not created in the galactic halos but, vise versa,
galaxies are formed around the (primordial) black hole seeds and not the mass of PBH is determined by the mass of the 
galaxy-host but the type and mass of the galaxy is determined by the original black hole seed.  New data reviewed below
strongly support this inverted picture.

The mass spectrum of the black holes observed in the Galaxy is also poorly understood, as well as the origin of MACHOs:
dim or absolutely non-luminous objects with masses about a half of the solar one, discovered through gravitational microlensing 
in the Galaxy and its halo. There are also quite a few not so striking but still significant problems, which do not fit the frameworks 
of the traditional cosmology and astrophysics.

The recent indications to the  black holes with intermediate masses of about 2000 solar mass inside the globular clusters  also
induced strong doubts on the canonic mechanism of such black hole creation.

The newest method of the determination of the stellar ages lead to an unexpected discovery of several "partiarches"  up to a one which
is older than the universe. Of course  the observation error cannot be excluded but anyhow such stars happen to be older that
the galaxy.

And last but not the least, the direct registration  of the gravitational waves by LIGO, which is reliably interpreted as a result of the
coalescence of the  black hole binary with masses of each companion of  about 30 $M_\odot$ and very low spins, can hardly 
be described by  the standard model of the black hole formation through the stellar collapse.

All mentioned above surprising discoveries (and many others) is very simply and naturally explained  by the single hypothesis of
the early creation of massive PBHs long before $z = 10$. Such a mechanism of PBH creation was suggested in ref.~\cite{AD-JS}.
In particular, the log-normal mass distribution of PBH was predicted there. This distribution became very popular during last year
in connection with the LIGO discovery and an understanding that the wide mass spectrum of PBH opens an intriguing 
possibility that PBH constitute the whole or a large fraction of the dark matter in the universe. This "new" idea was put
forward already in 1993 in the paper~\cite{AD-JS}, see also~\cite{DKK}.

The review is organized as follows. In the second section the observational data of the recent few years about the universe
at large redshifts,  $ z \sim 10$, are presented. In particular, the discoveries of the young bright galaxies, quasars (i.e. SMBH), supernovae,
gamma-bursters, and dust in the universe are discussed there. 
In the following  section the facts of life of the contemporary universe, mostly revealed during the observations of  the last decade, are presented.
All of them are not squeezed into the tight frameworks of the standard model. Section 4 is devoted to the formation and properties of the
black holes discovered by LIGO. The suggested solution of the related problems is based on the work~\cite{BDPP}.  
In sec. 5 the basic features of the model of the papers~\cite{AD-JS,DKK} of an early formation of PBHs and compact stellar-like objects 
is discussed and their mass spectrum is presented. In the next section 6 the evolution of the PBHs due to the  later matter accretion is  considered.
Section 7 is dedicated to the relation between the globular clusters and the intermediate mass black holes with  masses $\gtrsim 2000 M_\odot $
on the basis of the paper~\cite{AD-KP-gc}.  The impact of the early created compact astrophysical objects on the big bang nucleosynthesis
and on the spectrum of the angular fluctuations of CMB are briefly discussed in sec. 8. 
In sec. 9 a possibility of  the transition of the intensive gravitational waves into electromagnetic radiation in external magnetic field is considered.
In last section discussion and conclusion are presented.

In the review the results of the earlier conference talks are exploited~\cite{beasts}-\cite{talks-KAP}. On these talkes 
some more earlier references can be found.
\\[3mm]

The work was supported by the RSF Grant 16-12-10037.

\end{document}